# Quantum Amplitude Arithmetic

Shengbin Wang[1,*], Zhimin Wang[1,*], Guolong Cui[1], Lixin Fan[1], Shangshang Shi[1], Ruimin Shang[1], Wendong Li[1], Zhiqiang Wei[1,2,†] and Yongjian Gu[1,†]

[1] *College of Information Science and Engineering, Ocean University of China, Qingdao 266100, China*

[2] *High Performance Computing Center, Pilot National Laboratory for Marine Science and Technology (Qingdao), Qingdao 266100, China*

[*] *These authors contribute equally to this work.*

[†] *Correspondence author, e-mail: yjgu@ouc.edu.cn; weizhiqiang@ouc.edu.cn*

## ABSTRACT

Quantum algorithm involves the manipulation of amplitudes and computational basis, of which manipulating basis is largely a quantum analogue of classical computing that is always a major contributor to the complexity. In order to make full use of quantum mechanical speedup, more transformation should be implemented on amplitudes. Here we propose the notion of quantum amplitude arithmetic (QAA) that intent to evolve the quantum state by performing arithmetic operations on amplitude. Based on the basic design of multiplication and addition operations, QAA can be applied to solve the black-box quantum state preparation problem and the quantum linear system problem with fairly low complexity, and evaluate nonlinear functions on amplitudes directly. QAA is expected to find applications in a variety of quantum algorithms.

## I. INTRODUCTION

The quantum mechanical properties of superposition and entanglement underpin the promise of quantum computing that can provide exponential speedup over the classical computing [1]. Quantum computing brings about the mode of quantum parallelism for computation, which would be depicted by the quantum version of computing a function *f*, $U_f : \sum_x a_x |x\rangle |0\rangle \to \sum_x a_x |x\rangle |f(x)\rangle$. This transformation can be implemented by the quantum arithmetic on computational basis using the digital [2,3] or phase [4,5] methods. While quantum parallelism accomplishes nothing by itself, it typically works as the canonical starting point of quantum algorithms [6]. That is, the value of *f(x)* will then be transduced from the computational basis to the amplitude or phase for the further interference between states [7].

The above strategies are widely used in quantum algorithms, from the early Deutsch-Jozsa algorithm and Grover's search algorithm to the recent quantum linear system algorithm [8-10] and quantum deep learning algorithm [7,11,12]. However, since the cost of quantum arithmetic on the computational basis is high, the process of computing *f(x)* on basis and then transducing to amplitude would be a major contributor to the algorithm complexity. Considering the feature of NISQ devices [13], it is an interesting open problem to carry out such process more quantum mechanically.

In order to address this problem, here we propose a novel methodology called Quantum Amplitude Arithmetic (QAA). In contrast to the quantum arithmetic on basis or phase, QAA is designed to perform arithmetic on the complex amplitudes. By applying similar ideas, we have developed a fast quantum Poisson solver, which reduces the complexity overwhelmingly and is implementable on NISQ devices [14]. Here we extend such an idea to a complete framework of methodology.

## II. METHODS

The fundamental components of QAA are multiplication and addition operations on amplitudes. On the one hand, multiplication operation is straightforward because quantum computers use the architecture of product of unitary operation. The basic way of performing multiplication in QAA is shown in Fig. 1(a). On the other hand, addition operation on amplitudes is not natural for quantum computer because of the same reason of architecture. Fortunately, there exists a related technique in quantum algorithm, namely Linear Combination of Unitaries (LCU). LCU is a key subroutine of many important quantum algorithms, such as quantum linear systems algorithms [15] and Hamiltonian simulation algorithms [16,17]. Much inspired by the LCU method, we develop a way of linearly combining amplitudes, and this constitutes the addition operation of QAA as shown in Fig. 1(b). It is interesting to note that implementing linear combinations of unitaries is a fundamental operation for the duality quantum computer [18].

In general, multiplication and addition operations of QAA are implemented in a probabilistic way. The single qubit $R_y$ rotations are used to prepare the primitives of amplitude, i.e. $\cos\theta$ or $\sin\theta$ factor, and these are then used as blocks to construct the target amplitudes; that is, to perform arithmetic on the amplitudes.

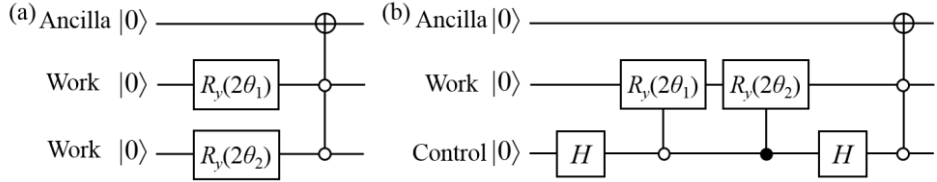

FIG. 1. The basic way of implementing (a) multiplication operation and (b) addition operation in QAA based on single qubit $R_y$ rotations. Just for the sake of illustration, one ancilla qubit and a Toffoli gate are used to present one possible results. Here the whole state can be expressed as (a) $(\cos\theta_1 \cdot \cos\theta_2)|001\rangle+|\omega\rangle$ and (b) $(\cos\theta_1 + \cos\theta_2)/2|001\rangle+|\omega'\rangle$ where $|\omega\rangle$ ($|\omega'\rangle$) is an unnormalized state containing the superposition state on the three qubits without $|001\rangle$. Further detailed discussion about (b) the addition operation is shown in appendix A.

## III. RESULTS

Most remarkably, the QAA method presented here offers an efficient solution for a wide class of important computational tasks. Below we show how QAA works on the fundamental problems including the black-box quantum state preparation problem and the quantum linear system problem, and finally we present a general way of evaluating nonlinear functions on amplitudes directly.

### A. Black-box quantum state preparation

Black-box quantum state preparation is an important subroutine in many quantum algorithms, which is leveraged to perform input preparation or state evolution. The first black-box state preparation algorithm was developed by Grover, and it requires to calculate arcsines on computational basis [19] that is a major contributor to the complexity. Most recently, Sanders, Low, Scherer and Berry proposed a new algorithm that avoids the need for arithmetic and reduce the complexity dramatically [20]. This work is inspirational in many ways [21] and suggests the significance of eliminating the basis arithmetic and conventional digit-analog conversion in quantum algorithm.

The central step of black-box state preparation is to implement such a transformation, $|x\rangle \mapsto \frac{x}{2^n}|x\rangle$ with $x$ being an $n$-bit integer. Note that an $n$-bit integer is itself a linear combination of binary bit as follows,

$$\frac{x}{2^n} = \sum_{i=1}^{n} \frac{1}{2^i} x_{n-i}, \qquad (1)$$

where $x_i$ is 0 or 1, and $x/2^n$ belongs to (0, 1). This equation can be implemented easily by QAA through adding all $1/2^i$ terms conditioned on the value of $x_i$. Specifically, scale up the quantum circuit of Fig. (1)b to one with $m = \lceil \log n \rceil$ control qubit; let each $R_y$ rotation be controlled by one bit of $x$ where the rotation angle is configured by $\theta_i =$ arcsin($1/2^i$). Fig. 2 shows the circuit representation of this algorithm.

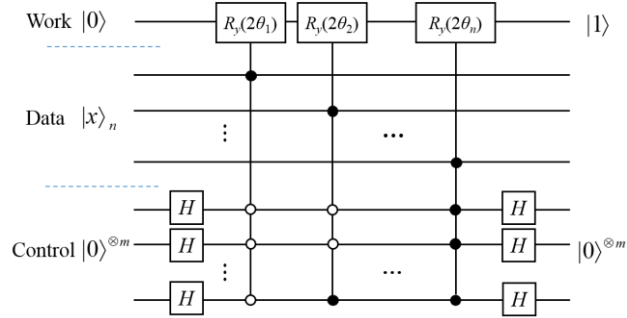

FIG. 2. The overall circuit for black-box state preparation based on QAA. The number of Control qubit is $m = \lceil \log n \rceil$ with $n$ being the number of Data qubit, so the number of extra qubits is log($n$)+1. The target state $\frac{x}{2^n}|x\rangle$ is indicated by $|0\rangle^{\otimes m}$ and $|1\rangle$ in registers Control and Work.

Through the circuit of Fig. 2, the quantum state evolves as follows,

$$|x\rangle_D |0\rangle_C^{\otimes m} |0\rangle_W \mapsto \frac{1}{2^m} \frac{x}{2^n} |x\rangle_D |0\rangle_C^{\otimes m} |1\rangle_W + |\omega\rangle_{D\otimes C\otimes W}, \qquad (2)$$

where $|\omega\rangle$ has no overlap on state $|0\rangle^{\otimes m}|1\rangle$ in registers Control and Work. The factor $1/2^m$, corresponding to the success probability of state preparation, can be further increased to 1/2 by combining the present circuit with the one discussed in detail in appendix B. More specifically, Eq. (1) could be re-written as

$$\frac{x}{2^{n+1}} = \frac{1}{2}\left[\sum_{i=1}^{m}\frac{1}{2^i}x_{n-i} + \sum_{k=m+1}^{n}\frac{1}{2^k}x_{n-k}\right], \tag{3}$$

where the first term of right hand side is implemented by the circuit in Fig. 7 of appendix B and the second term by the circuit of Fig. 2, which corresponds to part I and II in Fig. 3 respectively. With this improvement, the complexity of our algorithm for state preparation is $O(\log n)$ in extra qubits and $O(n\log n)$ in Toffoli gates, while as a contrast the complexity in Ref. [20] is $O(n)$ both in extra qubits and in Toffoli gates.

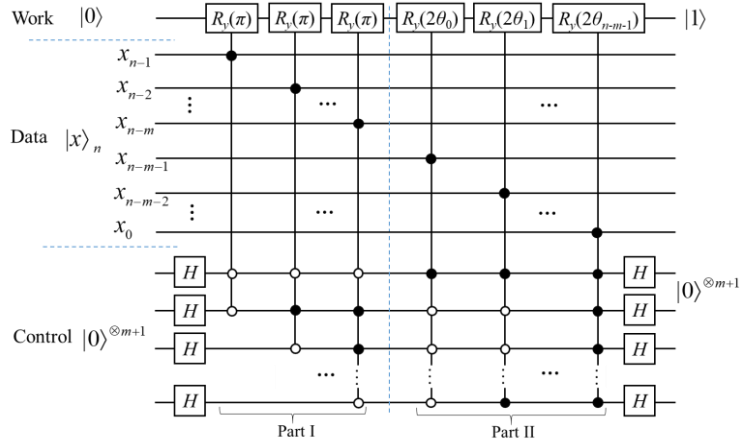

FIG. 3. The improved black-box quantum state preparation circuit. The circuit consists of two parts to implement respectively the first and second tem of right hand side of Eq. (3). The number of extra qubits is $m+2$, and the number of multi-controlled $R_y$ gates is $n$.

In order to prepare a complex amplitude, only one more extra qubit in the Control register is required. The real and imaginary part is prepared, respectively, by $R_y$ and $R_x$ gates controlled by the corresponding Data register. Additionally, if the new experimental techniques, multi-qubit controlled gates [22,23], are accessible, then the controlled rotations do not need to be decomposed as conventionally do [24] and the complexity of quantum operations becomes linear.

So far we show that QAA can be used to handle a problem associating with the linear combination of binary bit. Then it is natural to think of the Taylor series expansion that approximates a function by a linear combination of power functions. That combining QAA and Taylor expansion can result into an efficient algorithm for solving a particular class of quantum linear system problems.

## B. Quantum Linear System Problem

Quantum Linear System Problem is to solve a linear system of equations in a quantum way, that is, to prepare a quantum state $|x\rangle$ proportional to the solutions $\vec{x}$ of the linear system $A\vec{x}=\vec{b}$. Of which solving the tridiagonal linear system of equations is a fundamental computing task in many scientific and engineering applications [25]. Here we specifically discuss such a linear system that the coefficient matrix is an $N\times N$ tridiagonal Toeplitz matrix $A$ as follows,

$$A = \begin{pmatrix} 2y & -1 & & \\ -1 & 2y & & \\ & & \ddots & -1 \\ & & -1 & 2y \end{pmatrix}, \quad (4)$$

where $N = 2^n-1$. The eigenvalue is $\lambda_j = 2(y - \cos\frac{j\pi}{2^n})$ with $j = 1, 2 \ldots, N$ [26]. Note that when $y = 1$, it is actually the discretized matrix of one-dimensional Poisson equation discretized by the central difference method [27].

According to the spirit of the original Harrow-Hassidim-Lloyd algorithm [10], solving equations quantum mechanically is actually to prepare such a superposition states that the basis are the eigenstates of matrix $A$ and the amplitudes are the reciprocal of the corresponding eigenvalues. For the case of $y = 1$, as mentioned before, we have proposed an efficient method to prepare such superposition states [14]. Here for the general cases of $y \geq 2$, we utilize QAA and Taylor expansion to solve it.

The eigenvalue of matrix $A$ can be normalized to

$$x_j = 1 - \frac{\cos(j\pi/2^n)}{y}, \quad (5)$$

with $x_j = \lambda_j/2y$ belonging to $(1/2, 3/2)$. Then using the Taylor expansion, the reciprocal turns to

$$\frac{1}{x_j} = \prod_{i=0}^{m-1}[1 + (\frac{\cos(j\pi/2^n)}{y})^{2^i}] + \varepsilon, \quad (6)$$

where the truncation error satisfies $\varepsilon \leq 2^{-4n-4}$. The derivation of this equation is shown in appendix C.

The quantum circuit to implement Eq. (6) with $y = 2$ is shown in Fig. 4. In general, the circuit consists of two parts. The first part performs the multiplication between $\cos(\pi/3)$ (i.e. $1/y = 1/2$) and $\cos(j\pi/2^n)$ for $2^i$ times, and the second part implement the addition $1 + (\frac{1}{2}\cos\frac{j\pi}{2^n})^{2^i}$ and multiplication $\prod_i [1 + (\frac{1}{2}\cos\frac{j\pi}{2^n})^{2^i}]$. Note that this circuit is only for computing the reciprocal of eigenvalues; the complete circuit for solving the tridiagonal linear systems can be obtained by embedding this circuit into the algorithm framework in Ref. [14]. The total complexity of the present algorithm is $O(n)$ in qubits with a factor around 10 and $O(n^2)$ in one- and two-qubit quantum operations.

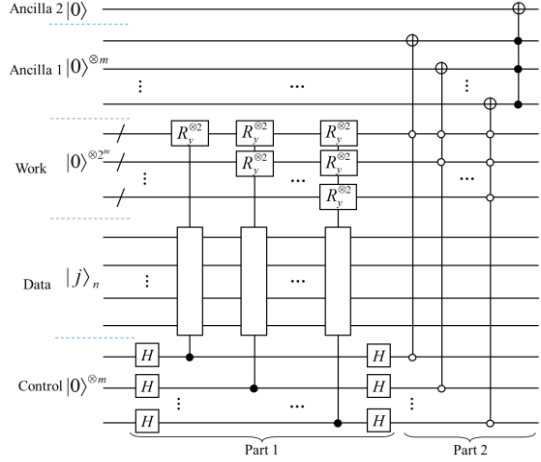

FIG. 4. The overall circuit for computing the reciprocal of eigenvalues according to Eq. (6) with $y = 2$. The circuit consists of two parts. Each $R_y^{\otimes 2}$ operator consists of one $R_y(2\pi/3)$ and one $R_y(j\pi/2^{n-1})$ rotation. The rectangle on the Data register represents that $R_y$ rotations are implemented conditioned on the value of $j$, and the specific circuit can be found in appendix A.

Taylor expansion could be used to approximate any function, but high degree of power functions are needed to reach required precision over a large interval and this would increase the complexity dramatically. In practical classical computing, piecewise polynomial approximation is a widely used method to evaluate functions with relatively low degree of polynomials [28].

### C. Piecewise polynomial approximation

Piecewise polynomial approximation is to partition the domain of a function into several parts and approximate each part by a linear combination of polynomials. Specifically, for a given function $f$ on an interval $[b_0, b_J]$, the domain is first divided into $[b_0, b_1], (b_1, b_2] \ldots (b_{J-1}, b_J]$, and in the $j^{th}$ subdomain the function is approximated by

$$p^{(j)}(x) = \sum_{i=0}^{d} p_i^{(j)} x^i, x \in (b_j, b_{j+1}] \qquad (7)$$

where $d$ is the degree of the polynomials and $p_i^{(j)}$ the coefficient of the $i^{th}$ term in the $j^{th}$ polynomial. The Remez algorithm [29] is used to obtain a satisfactory piecewise polynomial approximation for a given function and required precision; while the sparse-coefficient method [30] guarantees that the coefficient is accurately represented by $n$ bit, and it can also reduce the cost of data loading.

Now suppose that for a given function $f(x)$, all parameters $p_i^{(j)}$ have been obtained and stored in the QRAM [31]. The way of looping over all subdomains to evaluate the corresponding polynomial is similar as that in Ref. [2]; or the parallel polynomial evaluation scheme proposed there could be leveraged to parallelize the polynomial evaluation over the entire domain. Then QAA is applied to prepare a quantum state proportional to $\frac{p^{(j)}(x)}{2^n}|0\rangle$ as an approximation of the target state $\frac{f(x)}{2^n}|0\rangle$ in each subdomain. The algorithm consists of the following four steps, which is schematically shown in Fig. 5.

Step 1: in the Control register, allocate $m = \lceil \log(d+1) \rceil$ qubits and perform the transformation

$$|0\rangle^{\otimes m} \mapsto \frac{1}{\sqrt{d+1}} \sum_{i=0}^{d} |i\rangle. \tag{8}$$

Step 2: in the Parameter register, allocate $l = \lceil \log n \rceil$ qubits and invoke the parameters $p_i^{(j)}$ from the QRAM, then implement the following transformation using the same method as in the black-box state preparation (i.e. Eq. 3),

$$|i\rangle|0^l\rangle \mapsto |i\rangle(\frac{1}{2}\frac{p_i^{(j)}}{2^n}|0^l\rangle + |\omega\rangle_P). \tag{9}$$

Step 3: in the Data register, allocate $d$ qubits and compute $x^i$ on amplitude using the same method as in the quantum linear system problem (i.e. Eq. 6),

$$|i\rangle|0^d\rangle \mapsto |i\rangle(x^i|0^d\rangle + |\omega\rangle_D). \tag{10}$$

Now the entanglement state among the Control, Parameter and Data registers is

$$\frac{1}{\sqrt{d+1}} \sum_{i=0}^{d} |i\rangle(\frac{1}{2}\frac{p_i^{(j)} x^i}{2^n}|0^l\rangle|0^d\rangle + |\omega\rangle_{P\otimes D}). \tag{11}$$

Step 4: Finally, undo transformation in Control register, and the state evolves to

$$\frac{1}{2(d+1)}\frac{p^{(j)}(x)}{2^n}|0^m\rangle|0^l\rangle|0^d\rangle + |\omega\rangle_{C\otimes P\otimes D}. \tag{12}$$

Furthermore, amplitude amplification algorithm [32] can be applied to increase the success probability of obtaining the target state.

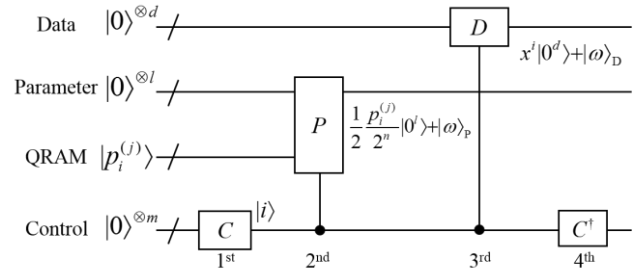

FIG. 5. The way of evaluating functions on amplitudes by combining QAA and piecewise polynomial approximation. The blocks $C$, $P$ and $D$ are used to implement the transformation of Eq. (8), (9) and (10), respectively. $C^\dagger$ represents the uncomputation of $C$.

## IV. CONCLUSIONS

In this paper, we propose the concept of Quantum Amplitude Arithmetic (QAA) that intends to evolve the quantum state by performing arithmetic operations on amplitude. We show that QAA can be used to solve the problem of black-box quantum state preparation efficiently using the fact that an integer is a linear combination of binary bit. By combining with Taylor series expansion, QAA can be used to solve the tridiagonal linear system of equations with very low complexity; and by combining with piecewise polynomial approximation, QAA can be used to evaluate nonlinear functions on amplitude directly. The advantage of Taylor expansion is that each series has the constant coefficient which simplifies the design of quantum circuit. The advantage of piecewise polynomial approximation is that it can approximate nonlinear functions efficiently with high accuracy, while the parameters of polynomials need to be computed at prior and stored in the QRAM.

It is expected that the present QAA method should be useful for many quantum problems, such as machine learning, pattern recognition, signal processing [10,33] etc. For example, one of the most important problems in the Quantum Convolutional Neural Network is to perform nonlinear activation functions, e.g. sigmoid, tanh and modified ReLU functions, on amplitude [34], which is also a common scenario in quantum machine learning. Our QAA method can be adapted to solve these problems immediately. In addition, there are other function approximation methods, such as argument reduction and look-up table [35], which can be explored to use in quantum computing in a similar way as we do here.

## ACKNOWLEDGEMENTS

This work was supported by the National Natural Science Foundation of China (Grants No. 61575180, 61701464), and the Pilot National Laboratory for Marine Science and Technology (Qingdao).

## APPENDIX A: ADDITION OF QAA

As a basic operation of QAA, the addition operation on amplitudes is designed based on the idea of Linear Combination of Unitaries (LCU). LCU provides a way of implementing a unitary by decomposing it into a linear combination of simple operators. Specifically, for a unitary $V$, it is decomposed into a linear combination of operator $U_i$ as $V = \sum_i p_i U_i$ with $p_i > 0$. Then define an operator $A$ that map state $|0^m\rangle$ to $\frac{1}{\sqrt{p}} \sum_i \sqrt{p_i} |i\rangle$ where $p = \sum_i p_i$, and an operator $U = \sum_i |i\rangle\langle i| \otimes U_i$. Thus, the operator $W = A^\dagger U A$ can evolve a state $|\psi\rangle$ as follows [16],

$$W|0^m\rangle|\psi\rangle = \frac{1}{p}|0^m\rangle V|\psi\rangle + |\Xi^\perp\rangle, \qquad (A1)$$

where the state $|\Xi^\perp\rangle$ is unnormalized and satisfies $(|0^m\rangle\langle 0^m|\otimes \mathbf{1})|\Xi^\perp\rangle = 0$.

Corresponding to the operators of LCU, the whole addition operation of QAA can be taken as the operator $W$, where the operator $A$ is the $H$ gates as shown in Fig.1(b) of the main paper and the operator $U$ is the two controlled $R_y$ rotations. According to Eq. (A1), the quantum state evolves through the circuit of Fig.1(b) as follows,

$$\begin{aligned}
W|0\rangle|0\rangle &= HUH|0\rangle|0\rangle = HU(\frac{1}{\sqrt{2}}(|0\rangle+|1\rangle)|0\rangle) \\
&= H\left[|0\rangle\langle 0|\otimes R_y(2\theta_1)+|1\rangle\langle 1|\otimes R_y(2\theta_2)\right](\frac{1}{\sqrt{2}}(|0\rangle+|1\rangle)|0\rangle) \\
&= \frac{1}{\sqrt{2}}H\left[|0\rangle R_y(2\theta_1)|0\rangle+|1\rangle R_y(2\theta_2)|0\rangle\right] \\
&= \frac{1}{2}\left[(|0\rangle+|1\rangle)(\cos\theta_1|0\rangle+\sin\theta_1|1\rangle)+(|0\rangle-|1\rangle)(\cos\theta_2|0\rangle+\sin\theta_2|1\rangle)\right] \\
&= \frac{1}{2}\begin{bmatrix}(\cos\theta_1+\cos\theta_2)|0\rangle|0\rangle+(\sin\theta_1+\sin\theta_2)|0\rangle|1\rangle \\ +(\cos\theta_1-\cos\theta_2)|1\rangle|0\rangle+(\sin\theta_1-\sin\theta_2)|1\rangle|1\rangle\end{bmatrix}, i.e. \begin{bmatrix}++ \\ --\end{bmatrix} \\
&= \frac{\cos\theta_1+\cos\theta_2}{2}|0\rangle|0\rangle+|\Xi^\perp\rangle
\end{aligned} \quad (A2)$$

The sum of $(\cos\theta_1+\cos\theta_2)$ is prepared on the amplitude of state $|00\rangle$.

For the implementation of $R_y(2\theta)$ rotations, $\theta$ is usually regarded as the control parameter. Suppose $\theta = j\pi/2^n$ and the binary representation of $j$ is $j = j_1 j_2 \cdots j_n = \sum_{k=1}^{n} 2^{n-k} j_k$, then the $R_y$ rotation can be expressed as

$$R_y(\frac{j\pi}{2^{n-1}}) = e^{-i\frac{j\pi}{2^n}Y} = e^{-i\frac{\pi}{2^n}(\sum_{k=1}^{n} 2^{n-k} j_k)Y} = \prod_{k=1}^{n} e^{(-i\frac{2^{n-k}\pi}{2^n}Y)\cdot j_k} = \prod_{k=1}^{n} R_y^{j_k}(\frac{\pi}{2^{k-1}}). \quad (A3)$$

Therefore, the two controlled $R_y$ rotations in Fig. 1(b) in the main paper can be implemented by the circuit as shown in Fig. 6. In addition, this circuit can also be used to implement the controlled $R_y$ rotations in Fig.3 in the main paper.

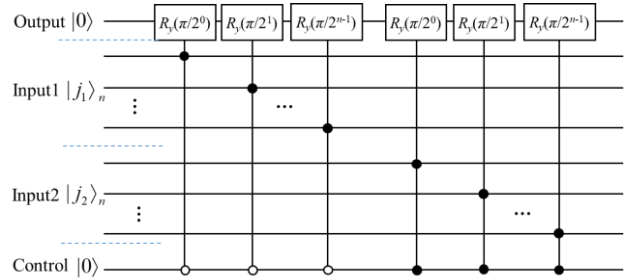

FIG. 6. The circuit to implement the two controlled $R_y$ rotations in the addition operation of QAA. The circuit complexity is $O(n)$.

## APPENDIX B: AN ALTERNATIVE WAY OF STATE PREPARATION

The binary expansion of an *n*-bit integer, i.e. Eq. (1) in the main paper, could be rewritten as

$$\frac{x}{2^n} = \sum_{i=1}^{n} 2^{n-i} \frac{1}{2^n} x_{n-i}. \tag{B1}$$

It shows that the factor of each $x_{n-i}$ is $2^{n-i}/2^n$, which can be interpreted as that the factor of $x_{n-i}$ is $1/2^n$ with a number of $2^{n-i}$. Therefore, the right hand side of Eq. (B1) could be regarded as a sum of $2^n-1$ terms, and each of which has a factor of $1/2^n$, then Eq. (B1) turns to be

$$\frac{x}{2^n} = \sum_{i=1}^{n} \sum_{k=1}^{2^{n-i}} \frac{1}{2^n} x_{n-i}. \tag{B2}$$

Based on the principle of QAA, one natural way of implementing Eq. (B2) is that first generate $2^n$ uniform superposition states in the control register and then prepare the $2^n-1$ terms of Eq. (B2) conditioned by the control qubits and the value of $x_{n-i}$. Obviously, operations on the terms corresponding to the same $x_{n-i}$ can be combined together. Specifically, since the number of term $x_{n-1}$ is $2^{n-1}$, half of the $2^n$ uniform superposition states are entangled with $x_{n-1}$ bit, namely those bases in Control register with highest bit being 0 as shown in Fig. 7. Similarly, the number of term $x_{n-2}$ is $2^{n-2}$, so half of the remaining $2^{n-1}$ superposition states, namely the control states with highest bit being 1 and second being 0, are entangled with $x_{n-2}$. The same is true for $x_{n-3}$ and so on.

The complexity of the algorithm is $O(n)$ in extra qubits and $O(n^2)$ in Toffoli gates, which is worse than that of Ref. [20]. However, by combing the present one with the method in the main paper, the complexity can be reduced to $O(\log n)$ in extra qubits and $O(n\log n)$ in Toffoli gates as discussed in the main paper.

In Ref. [20], $|x\rangle$ is compared with the uniform superposition states in an *n*-qubit control register, so the number of basis whose value is less than that of $|x\rangle$ is just $x$ itself, which also aim at finding the $x$ basis from the $2^n$ basis. From this point of view, the method presented here is similar to Ref. [20].

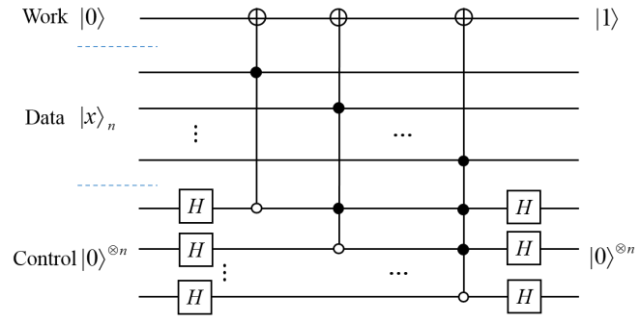

FIG. 7. An alternative way of implementing Eq. (1) in the main paper.

As can be seen from the above discussion, black-box state preparation circuit can be implemented in different ways because the binary expansion of an integer can be expressed in different forms, such as Eqs. (1) and (3) in the main paper and Eq. (B2). This implies the tremendous potential of applications of QAA.

# APPENDIX C: RECIPROCAL OF EIGENVALUES

Assume the Taylor series $1/x = \sum_{i=0}^{\infty}(1-x)^i$ is truncated at the $k^{\text{th}}$ term $(1-x)^{k-1}$, then the truncation error is

$$\begin{aligned}\varepsilon &= \sum_{i=k}^{\infty}(1-x)^i = (1-x)^k + (1-x)^{k+1} + \cdots \\ &= (1-x)^k[1+(1-x)+(1-x)^2+\cdots] \\ &\leq \frac{(1-x)^k}{x} \leq \frac{(1-1/2)^k}{1/2} \\ &= 2^{-k+1}\end{aligned} \quad (C1)$$

In order to reach a precision of $\varepsilon \leq 2^{-4n-4}$ for the eigenvalues scaled into (1/2, 3/2) as shown in Eq. (5) in the main paper, at least $4n+5$ terms are needed. Note that the precision increases exponentially while the number of qubit increases linearly.

We can, of course, calculate these $4n+5$ terms separately and then add them together by QAA, but the complexity is still a little high. To reduce the complexity, the following trick can be used to convert the sum to a product when the number of truncated terms equals to $2^m$. Assume $k = 2^m$, then we have

$$\begin{aligned}\frac{1}{x} &= 1+(1-x)+\cdots+(1-x)^{k-1}+\varepsilon \\ &= [1+(1-x)+\cdots+(1-x)^{k/2-1}]+(1-x)^{k/2}[1+(1-x)+\cdots+(1-x)^{k/2-1}]+\varepsilon \\ &= [1+(1-x)^{k/2}][1+(1-x)+\cdots+(1-x)^{k/2-1}]+\varepsilon \\ &= \cdots \\ &= [1+(1-x)^{2^{m-1}}][1+(1-x)^{2^{m-2}}]\cdots[1+(1-x)^{2^0}]+\varepsilon\end{aligned} \quad (C2)$$

Substitute Eq. (5) in the main paper into Eq. (C2), then we have

$$\frac{1}{x_j} = [1+(\frac{\cos(j\pi/2^n)}{y})^{2^{m-1}}][1+(\frac{\cos(j\pi/2^n)}{y})^{2^{m-2}}]\cdots[1+(\frac{\cos(j\pi/2^n)}{y})^{2^0}]+\varepsilon. \quad (C3)$$

And finally we have

$$\frac{1}{\lambda_j} = \frac{1}{2y}[1+(\frac{\cos(j\pi/2^n)}{y})^{2^{m-1}}][1+(\frac{\cos(j\pi/2^n)}{y})^{2^{m-2}}]\cdots[1+(\frac{\cos(j\pi/2^n)}{y})^{2^0}]+\varepsilon', \quad (C4)$$

where $\varepsilon' \leq (1/y)2^{-4n-5}$. With this equation, the number of qubit in Control register in Fig. 4 of the main paper is $m = \lceil \log(4n+6+\log y) \rceil$. Comparing with the way of addition as shown in Eq. (C2), the circuit complexity can be reduced one order both in qubits and in gates. For instance, when $y = 2$, we choose $m = \lceil \log(4n+7) \rceil$ and Eq. (C3) turns to

$$\frac{4}{\lambda_j} = [1+(\frac{1}{2}\cos\frac{j\pi}{2^n})^{2^{m-1}}][1+(\frac{1}{2}\cos\frac{j\pi}{2^n})^{2^{m-2}}]\cdots[1+(\frac{1}{2}\cos\frac{j\pi}{2^n})^{2^0}]+\varepsilon, \quad (C5)$$

where the factor 1/2 can be taken as $\cos(\pi/3)$.